\documentclass[twoside,twocolumn,9pt]{article}
\usepackage{extsizes}
\usepackage[super,sort&compress,comma]{natbib} 
\usepackage[version=3]{mhchem}
\usepackage[left=1.5cm, right=1.5cm, top=1.785cm, bottom=2.0cm]{geometry}
\usepackage{balance}
\usepackage{times,mathptmx}
\usepackage{sectsty}
\usepackage{graphicx} 
\usepackage{lastpage}
\usepackage[format=plain,justification=justified,singlelinecheck=false,font={stretch=1.125,small,sf},labelfont=bf,labelsep=space]{caption}
\usepackage{float}
\usepackage{fancyhdr}
\usepackage{fnpos}
\usepackage[english]{babel}
\usepackage{array}
\usepackage{droidsans}
\usepackage{charter}
\usepackage[T1]{fontenc}
\usepackage[usenames,dvipsnames]{xcolor}
\usepackage{setspace}
\usepackage[compact]{titlesec}
\usepackage{upgreek}

\definecolor{cream}{RGB}{222,217,201}

\begin{document}

\pagestyle{fancy}
\thispagestyle{plain}
\fancypagestyle{plain}{

}

\makeFNbottom
\makeatletter
\renewcommand\LARGE{\@setfontsize\LARGE{15pt}{17}}
\renewcommand\Large{\@setfontsize\Large{12pt}{14}}
\renewcommand\large{\@setfontsize\large{10pt}{12}}
\renewcommand\footnotesize{\@setfontsize\footnotesize{7pt}{10}}
\makeatother

\renewcommand{\thefootnote}{\fnsymbol{footnote}}
\renewcommand\footnoterule{\vspace*{1pt}%
\color{cream}\hrule width 3.5in height 0.4pt \color{black}\vspace*{5pt}} 
\setcounter{secnumdepth}{5}

\makeatletter 
\renewcommand\@biblabel[1]{#1}            
\renewcommand\@makefntext[1]%
{\noindent\makebox[0pt][r]{\@thefnmark\,}#1}
\makeatother 
\renewcommand{\figurename}{\small{Fig.}~}
\sectionfont{\sffamily\Large}
\subsectionfont{\normalsize}
\subsubsectionfont{\bf}
\setstretch{1.125} 
\setlength{\skip\footins}{0.8cm}
\setlength{\footnotesep}{0.25cm}
\setlength{\jot}{10pt}
\titlespacing*{\section}{0pt}{4pt}{4pt}
\titlespacing*{\subsection}{0pt}{15pt}{1pt}

\fancyfoot{}
\fancyhead{}
\renewcommand{\headrulewidth}{0pt} 
\renewcommand{\footrulewidth}{0pt}
\setlength{\arrayrulewidth}{1pt}
\setlength{\columnsep}{6.5mm}
\setlength\bibsep{1pt}

\makeatletter 
\newlength{\figrulesep} 
\setlength{\figrulesep}{0.5\textfloatsep} 

\newcommand{\topfigrule}{\vspace*{-1pt}%
\noindent{\color{cream}\rule[-\figrulesep]{\columnwidth}{1.5pt}} }

\newcommand{\botfigrule}{\vspace*{-2pt}%
\noindent{\color{cream}\rule[\figrulesep]{\columnwidth}{1.5pt}} }

\newcommand{\dblfigrule}{\vspace*{-1pt}%
\noindent{\color{cream}\rule[-\figrulesep]{\textwidth}{1.5pt}} }

\makeatother

\twocolumn[
  \begin{@twocolumnfalse}
\vspace{3cm}
\sffamily
\begin{tabular}{m{4.5cm} p{13.5cm} }

{ } & \noindent\LARGE{\textbf{Effective energy density determines the dynamics of suspensions of active and passive matter}} \\

\vspace{0.3cm} & \vspace{0.3cm} \\

 & \noindent\large{Ryan Krafnick\textit{$^{a}$} and Angel E. Garc\'{i}a\textit{$^{ab}$}} \\

\vspace{0.3cm} & \vspace{0.3cm} \\
{\today} & \noindent\normalsize{The unique properties of suspensions containing both active (self-propelling) and passive matter, arising from the nonequilibrium nature of these systems, have been widely studied (e.g., enhanced diffusion, phase separation, and directed motion).  Despite this, our understanding of the specific roles played by the relevant parameters of the constituent particles remains incomplete.  For instance, to what extent are the velocity and density of swimmers qualitatively distinguishable when it comes to the resultant properties of the suspension as a whole, and when are they merely two different realizations of the same thing?  Through the use of numerical simulations, containing both steric and hydrodynamic interactions, we investigate a new parameter, the effective energy density, and its ability to uniquely describe the dynamics and properties of a hybrid system of active and passive particles, including the rate of pair formation and the energy distribution amongst different constituent elements.  This parameter depends on both the density and the swimming velocity of the active elements, unifying them into a single variable that surpasses the descriptive ability of either alone.} \\

\end{tabular}

 \end{@twocolumnfalse} \vspace{0.6cm}

  ]

\renewcommand*\rmdefault{bch}\normalfont\upshape
\rmfamily
\section*{}
\vspace{-1cm}


\footnotetext{\textit{$^{a}$Department of Physics, Applied Physics, and Astronomy and The Center for Biotechnology and Interdisciplinary Studies, Rensselaer Polytechnic Institute, Troy, New York 12180, USA.}}
\footnotetext{\textit{$^{b}$Center for Nonlinear Studies, CNLS MS B258, Los Alamos National Laboratory, Los Alamos, New Mexico 87545, USA. E-mail: agarcia@lanl.gov}}



\section{Introduction}
Active matter denotes a class of both living and non-living matter which has the capacity to self-propel, through the use of either external (e.g., laser light \cite{Wurger2014}) or internal (e.g., cell metabolism \cite{Shapiro1995}) energy sources.  The methods of propulsion are varied, from the use of flagella at the microscale, as in \emph{Chlamydomonas reinhardtii},\cite{Tuval2009} to self-phoresis, as utilized by Janus particles,\cite{Zhao2011,Sen2012} and macroscopic beating wings, as in bird flight.\cite{Warrick1997} While the mechanics of individual swimmers are interesting to investigate in their own right,\cite{Zhang2012} collections of active matter additionally exhibit a variety of unique, emergent properties, including self-assembly,\cite{Cacciuto2016} collective motion,\cite{Kessler2007,Goldstein2007,Schmidt2013,Goldstein2014,Goldstein2007b}, and clustering.\cite{Speck2013,Stark2014,Lowen2015,Pedley2008} These properties come from the ability of active matter to drive itself out of thermal equilibrium,\cite{Ramaswamy2010} violating the fluctuation dissipation theorem (FDT),\cite{Yodh2007} although the applicability of equilibrium concepts and the degree of the departure remain in question.\cite{vanWijland2016,Brady2015,Hagan2016} The passive (non-self-propelling) particles in \emph{hybrid} systems, which contain both passive and active matter, also gain new properties owing to their swimming neighbors.  For example, passive particle diffusion significantly increases,\cite{Marenduzzo2011,Arratia2016,Graham2009,Libchaber2000,Goldstein2009,Graham2008} and effective attractive interactions develop between otherwise noninteracting particles.\cite{DiLeonardo2011,Garcia2015} The applications of active matter, beyond supplying information about fundamental, ubiquitous systems, include drug and cargo delivery,\cite{Sen2013} micromotors,\cite{DiLeonardo2015,Ruocco2009} water remediation,\cite{Schmidt2013b} and biosensing.\cite{Wilson2014} A thorough review of the past, present, and future of the field of active matter has recently been published.\cite{Volpe2016}

An understanding of how the aforementioned properties depend upon the parameters of a given system is, naturally, of fundamental importance, and much progress has already been made to this end.  For example, one of the primary targets of research has been quantifying the enhancement of particle diffusion, which has been shown to be non-monotonic in passive particle size\cite{Arratia2016} and increases approximately linearly with swimmer density,\cite{Libchaber2000,Graham2009} as does the mean squared velocity.\cite{Graham2009} In addition, raising the swimmer rotational diffusivity (alternatively, lowering the correlation time or velocity) has been shown to enhance the stability of solutions\cite{Underhill2013} and reduce the rotational diffusivity of passive tracers.\cite{Wu2014} Many of the \emph{effects} of swimmer density and velocity are, however, coupled, and cannot be separated, except under certain simplifying assumptions (e.g., no interactions between swimmers).  This is due to the hydrodynamic interactions between swimmers,\cite{Kessler2007,Subramanian2010} which cannot be qualitatively mimicked by steric effects,\cite{Garcia2015} and depends on the specific nature of the active matter itself.

Given this realization, we propose a new parameter, called the effective energy density, which combines the swimmer density and velocity by accounting for the system-dependent coupling.  The \emph{input} energy density of a system is $\epsilon = \rho v^2$, where $\rho$ is the number density of swimmers and $v$ is the swimming velocity.  However, the effective (as experienced by the system) energy density, $\epsilon^*$, would only equal $\epsilon$ under the condition of completely independent swimmers.  In reality, swimmers may preferentially align themselves in parallel or perpendicular (or intermediate angles), in pairs or higher order structures, depending on the specific nature of the swimming mechanism and the shape of the swimmer.\cite{Simha2013}  This causes the measured energy density to change.  For example, consider a suspension that has doubled the number of swimmers, but each swimmer pairs up with one of the original swimmers.  This is effectively an unchanged number density, with double the velocity (since the hydrodynamic flow fields superpose), which yields $\epsilon^*_2 = 4\epsilon_1$; rather than double the energy density, we have quadruple.  While this example is extreme, it illustrates the impact that interactions within a system cause, such that density and velocity (as \emph{measured} by other particles within the solution) are dependent upon one another.  We test the efficacy of this parameter by analyzing a series of simulations, each with different values of the density and velocity.  In particular, we find that it excels at predicting the distribution of energy within a solution, the strength of transient vortices, and the rate of passive particle interactions, in addition to clarifying a discrepancy in the density dependence of passive particle diffusivity.

We perform simulations on a suspension containing both passive spheres and active rods, using a lattice Boltzmann fluid.\cite{Succi2001}  The rods execute run-and-tumble motion, which involves alternating periods of constant-velocity, one-directional propulsion and randomized reorientation.  This method of movement is used by a variety of microscopic life forms, including \emph{Escherichia coli}, and allows an organism to perform a random walk through its environment at long time scales (which, \emph{in natura}, is biased toward preferable environmental conditions\cite{Brown1972}).  Run-and-tumble motion can be mapped to active Brownian motion,\cite{Tailleur2013} wherein the reorientation is driven by ongoing fluctuations, rather than separate phases, and includes, for example, the dynamics of Janus spheres.\cite{Wurger2014,Speck2013,Zhao2011}  The rods are so-called pushers, which swim by pushing against the fluid behind them (when combined with the viscous friction force of the body on the fluid, the motion is force-free).  This is in contrast to pullers, which pull on the fluid in front of them in order to achieve movement, and include \emph{C. reinhardtii} (which is also a run-and-tumble particle\cite{Goldstein2009b}). Another common variety is the squirmer, which moves via a prescriped tangential surface velocity,\cite{Pedley2006} and includes \emph{Paramecium caudatum}.\cite{Hota2006} There are additionally variants of active matter that contribute to a system via rotation rather than translation.\cite{AlexanderKatz2016,Glotzer2015} Each method of swimming yields different hydrodynamic interactions between swimmers and passive particles, and will necessarily result in a different coupling between density and velocity, but the concept of effective energy density can be applied to all of these.

\section{Method}
Simulations are performed in a two-dimensional (2D) box of size $L$, similar to the quasi-2D format of certain experiments.\cite{Libchaber2000,DiLeonardo2011} Passive particles are represented as discs of diameter $d$, and active particles are rods of length $l$ and width $w$.  For these simulations, $L/d \approx 200$ and the boundary conditions are periodic.  The aspect ratio of the swimmers is $l/w=3$, and, for the size of \emph{E. coli} ($w$ = 1 $\upmu$m), the simulations equate to roughly 1 minute in a 1 mm\textsuperscript{2} film, with a 1/6 $\upmu$s time step.  Hydrodynamic interactions and flow fields are evaluated through the use of the Lattice Boltzmann Method,\cite{Succi2001} which represents the fluid velocity distribution at every point on a lattice and evolves through alternating collision and propagation phases.  The fluid-solid coupling follows the immersed boundary method,\cite{Peskin2002} and the cell size for this work is set equal to $w$. The interaction between an object and the fluid is calculated with a friction force,
\begin{equation}
 \mathbf{F}_f(\mathbf{r}_i) = \gamma_i\sum_j(\mathbf{u}(\mathbf{r}_i^j) - \mathbf{v}_i^j),
\end{equation}
where $\mathbf{u}$ is the fluid velocity, $\mathbf{v}_i^j$ is the particle velocity at node $j$, and $\mathbf{r}_i^j$ is the position of the $j$th node along the surface of the object, with nodes placed roughly $w$ apart from one another; the force is calculated as a sum over $j$.  The coupling constant, $\gamma_i$, is proportional to the density and kinematic viscosity of the fluid, along with a factor determined by the shape of a given particle.  This is simply another way of writing Stokes' Law, $F_d = 6 \pi \mu R V$, where $6 \pi R$ is the shape factor (for a sphere), $\mu$ is the dynamic viscosity (density times kinematic viscosity), and $V$ is the relative velocity of the fluid with respect to the embedded object (represented here as $\mathbf{u} - \mathbf{v}$).

Simulations take place in the low Reynolds number ($Re$) regime, $Re=vl/\nu \ll 1$, where $v$ is the particle velocity, $l$ is the particle length, and $\nu$ is the fluid kinematic viscosity.  For these simulations, the largest value of $Re$ is $9 \times 10^{-5}$.  $Re$ represents the ratio between inertial and viscous forces; a low value indicates very little system memory, where the friction eliminates momentum against the current.  Here the fluid velocity experiences only transient departures from the particle velocity along a particle's surface, and the particle velocity essentially reduces to $\mathbf{v}=\mathbf{F}/\gamma$.

Swimmers experience an additional interaction with the fluid, owing to their self-propulsion and tumbling phases, given by
\begin{align}
 \begin{split}
  \mathbf{F}_r &= \zeta \mathbf{\hat{e}}(1 - \theta_i), \\
  \mathbf{T}_r &= \xi_i(t) \theta_i,
 \end{split}
\end{align}
where $\zeta$ is a constant that determines the swimming speed ($v = \zeta / \gamma$), $\xi_i(t)$ supplies Gaussian distributed random values, $\mathbf{\hat{e}}$ is a unit vector pointing in the direction a swimmer faces, and $\theta_i$ is a boolean indicating the phase ($\theta_i=0$ for run, $\theta_i=1$ for tumble).  All fluid interactions have an equal and opposite force applied onto the fluid itself, and the motion is force-free.

Steric effects are represented by an excluded volume repulsive force directed perpendicular to the contacting surfaces, given by
\begin{equation}
 \mathbf{F}_e(\mathbf{r}_i,\mathbf{r}_j) \propto max(a_i + a_j - |\mathbf{r}_i-\mathbf{r}_j|, 0),
\end{equation}
where $a_i$ is the radius ($d/2$ or $w/2$) of the $i$th particle, and $\mathbf{r}_i$ is the position of the $i$th particle.  For discs, $r$ is at the center, and for rods, $r$ is the point along the centerline closest to the point of overlap.  The position and angle of each particle are updated at each time step by integrating the force and torque using a leapfrog algorithm.  In order to focus on the effect of the active matter itself, we have neglected thermal fluctuations, which, in practice, have a considerably smaller effect.\cite{Marenduzzo2011}  For more detail, please see our previous work.\cite{Garcia2015}

The primary simulation for this work uses the parameters of our earlier work, which are set to give reasonable conditions for a system of \emph{E. coli} swimmers and latex beads.  That is, for a length scale of $w$ = 1 $\upmu$m, the kinematic viscosity and fluid density match water, and the swimming velocity is 30 $\upmu$m/s, suitable for \emph{E. coli}.  The ratio of swimmers to passive particles is 10:1, with a total area fraction of 0.1 and a swimmer area fraction of $\phi = 0.06$.  The 15 additional simulations reduce either the density or the velocity of the swimmers in order to probe completely separate locations of parameter space, with a low of 1/16 times the primary values.

In order to calculate the effective energy density $\epsilon^*$ of a simulation, the swimmers must be partitioned into clusters, by use of a clustering algorithm.  In the present case - rod-like, run-and-tumble pushers - the relevant organization is adjacent and parallel.  Fluid entrainment inwards from both sides causes parallel swimmers to attract one another, and the orientation (parallel or antiparallel) does not change the shape of the flow, since the objects in question create symmetric, outward flows from the front and the back.  Thus, the important criteria for two objects to be considered in the same cluster is the distance between their centers.  A cutoff of $1.5w$ was used for clustering, and a single cluster accounts for all connected swimmers; i.e., cluster size is not the number of particles within $1.5w$ of an individual swimmer, but rather the number of particles that can be connected through a path of $1.5w$ or less distances between their centers.  This cutoff equates to a perpendicular gap of only half the swimmer width, or a parallel separation of about one width, for example.

We approximate the velocity of a cluster of size $N$ as $Nv$, where $v$ is the swimming speed of an independent swimmer.  As a cluster, or raft, of swimmers grows in number, this approximation becomes less accurate, owing to the spacial scale of the unit, but the majority of clusters are two or fewer in size, so this approximation is satisfactory.  For a number density $\rho_i$ of clusters of size $N_i$, the partial energy density contribution is thus $\epsilon_i^* = \rho_i(N_iv)^2 / \epsilon$.  The total effective energy density is thus
\begin{equation}
 \epsilon^* = \sum_{i=1}^\infty \rho_i(N_iv)^2 / \epsilon.
\end{equation}
Here a value of $\epsilon^* = 1$ (which is the minimum) indicates that the effective energy density is unchanged from the presupposed value, and necessarily requires all swimming units to be of size one.  While the sum theoretically extends to infinity, the largest measured cluster gave $\rho_9=4.7 \times 10^{-8}$, which equates to an immeasurable effect $\epsilon_9^*=3.8 \times 10^{-6}$.  We note that the present values of active matter density are well below the threshold of large-scale collective behavior.  In such a limit of high density, more important elements than parallel groups of swimmers would be needed to accurately describe $\epsilon^*$.  For example, this simple clustering algorithm couldn't possibly incorporate the dynamics of bacterial turbulence\cite{Goldstein2013} or the Zooming BioNematic.\cite{Kessler2007}

\section{Results}
\begin{figure}
\centering
  \includegraphics[width=\linewidth]{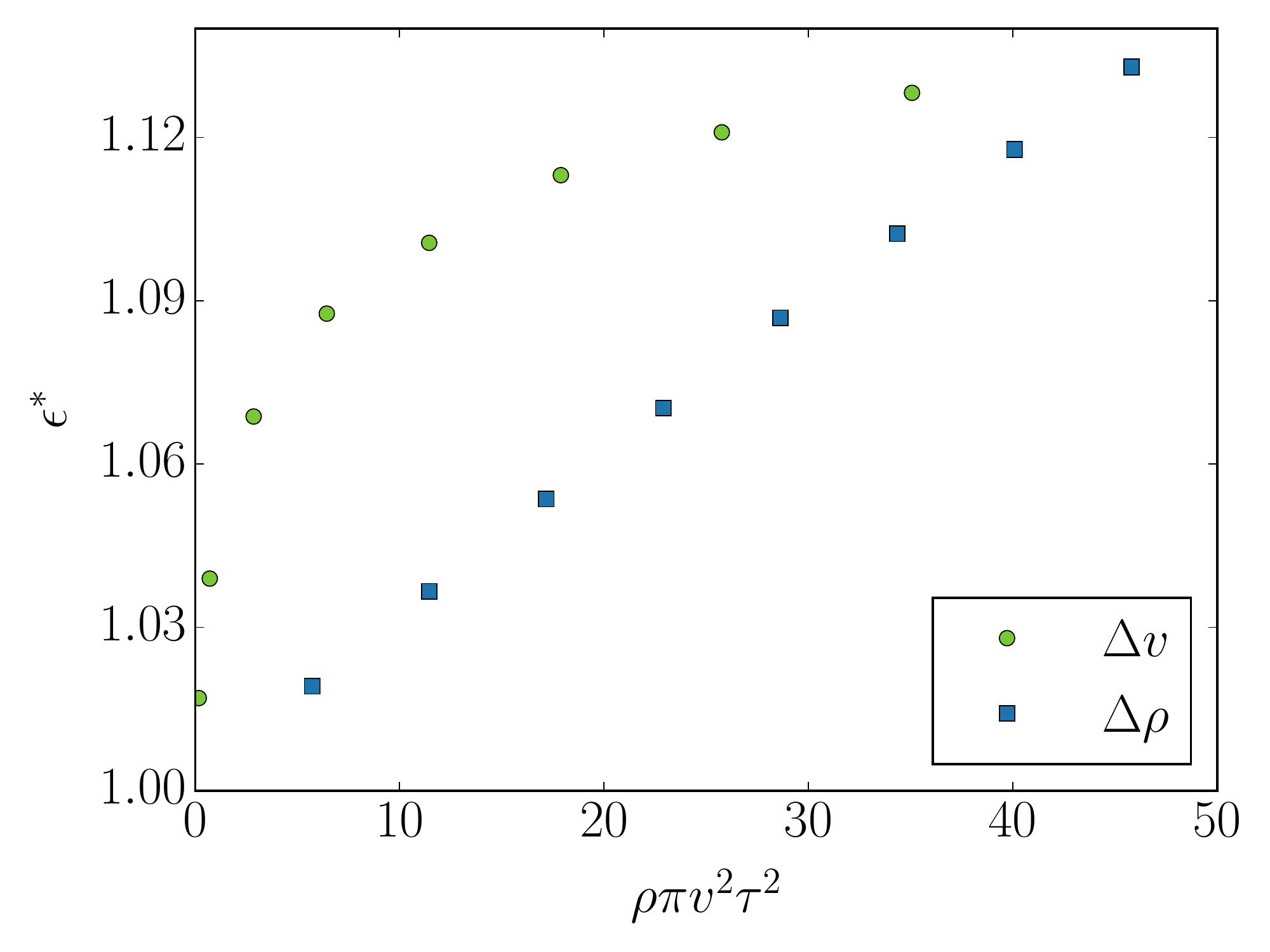}
  \caption{Effective energy density $\epsilon^*$ as a function of the unitless input energy density $\epsilon$, where $\rho$ is the swimmer number density, $v$ is the swimmer run velocity, and $\tau$ is the run (correlation) time.  The x-axis thus represents the number of other swimmers that a swimmer can reach during its run phase.  Green circles indicate constant density with varying velocity values, and blue squares indicate constant velocity with varying density values.  Each parameter causes a different reaction in $\epsilon^*$, and this depends on the shape and nature of the swimmers themselves, as well as the fluid.}
  \label{fgr:energy_density}
\end{figure}

The effective energy density $\epsilon^*$, as calculated by the aforementioned clustering algorithm, is shown in Fig. \ref{fgr:energy_density}, as a function of the (unitless) input energy density $\epsilon = \rho \pi v^2\tau^2$.
This is equivalent to the typical number of swimmers within a circle of radius $v\tau$ equal to the average distance traveled during a run phase.  This plot is not universal, as the shape will necessarily depend on the physical nature of the swimmers and the fluid itself.  Number density dependence is roughly linear, since the likelihood of meeting up with another swimmer and forming a pair is proportional to $\rho$; if there are twice as many particles, there are twice as many collisions (per particle), which leads to double the departure from $\epsilon$.  In contrast, the (squared) velocity dependence appears to saturate relatively quickly.  This is related to the fixed value of the correlation time $\tau$, which is the typical duration of a swimmer's run phase.  At low velocities, the swimmers do not have enough time to reach each other before changing course, at which point any favorable orientations will most likely be lost.  Thus, more consecutive tumbling phases must result in adequate alignment, reducing the chance of ever forming a cluster.  Keep in mind that, during tumbling, the effective radius of a swimmer is equal to $l/2$, so that the minimum separation is no longer $w$.  Values of $\epsilon < 1$ do not give enough time, on average, for nearby particles to meet, since this is the number of particles within a swimmer's run distance.  In fact, if we assume a correlation length of $\epsilon = 1$ (for an exponential decay), the estimated asymptotic value is ${\approx}1.14$, which is a good fit to the data.  Clearly, $\epsilon$ and $\epsilon^*$ yield completely different maps of $\rho$ and $v$ onto the same axis; therefore, whenever one of them adequately describes the data, the other necessarily fails.  We will show that $\epsilon^*$ does an excellent job in unifying the parameters investigated in this work.

\begin{figure}
\centering
  \includegraphics[width=\linewidth]{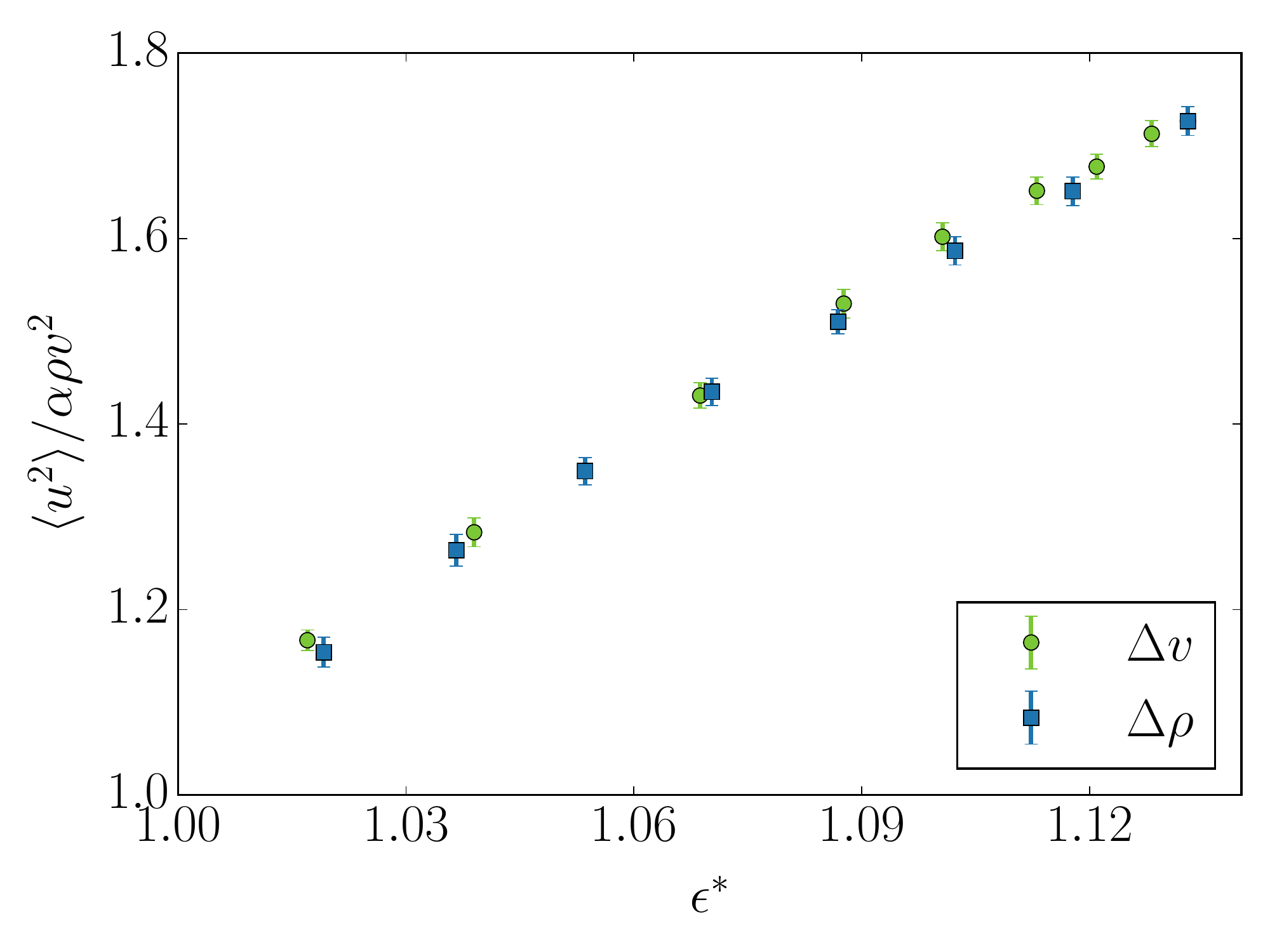}
  \caption{Fluid energy density $\langle u^2 \rangle$ as a function of the effective energy density $\epsilon^*$, where $\rho$ is the swimmer number density, $v$ is the swimmer run velocity, and $\alpha$ is the fractional energy density of an isolated swimmer multiplied by the domain size $L^2$.  The ideal limit would be a constant value of 1.  Green circles indicate constant density with varying velocity values, and blue squares indicate constant velocity with varying density values.  $\epsilon^*$ uniquely captures the fluid energy, despite the two curves taking paths through separate parameters.}
  \label{fgr:location_energy}
\end{figure}

As a first test, we examine the energy contained in the fluid itself, as represented by the average square fluid velocity, $\langle u^2 \rangle$, as a function of $\epsilon^*$, in Fig. \ref{fgr:location_energy}.
The constant $\alpha$ is the fractional energy an isolated swimmer gives to the fluid, multiplied by the size of the domain, $L^2$; this constant normalizes the y-axis to a minimum value of 1 and makes the quantity unitless.  As can be seen, the fluid reaches this theoretical value only at the limit of zero swimmer density and velocity, and otherwise extends well above it.  Furthermore, the curves for varying $\rho$ and $v$ show clear alignment owing to the effective energy density construct.  In other words, the energy contained in the fluid \emph{is} properly described by the simple clustering algorithm, and cannot be explained by assuming independent swimmers operating at the input density and velocity.  The departure of $\langle u^2 \rangle$ from the ideal value is more pronounced than that of $\epsilon^*$, owing to coordination between swimming units at scales larger than the size of a swimmer (which is the limitation of the clustering algorithm).  While, as mentioned above, we are not simulating systems in the limit of collective motion, patterns do still emerge, which will be examined below.  Surprisingly though, our naive approximation of $\epsilon^*$ still uniquely determines the value of $\langle u^2 \rangle$.

\begin{figure}
\centering
  \includegraphics[width=\linewidth]{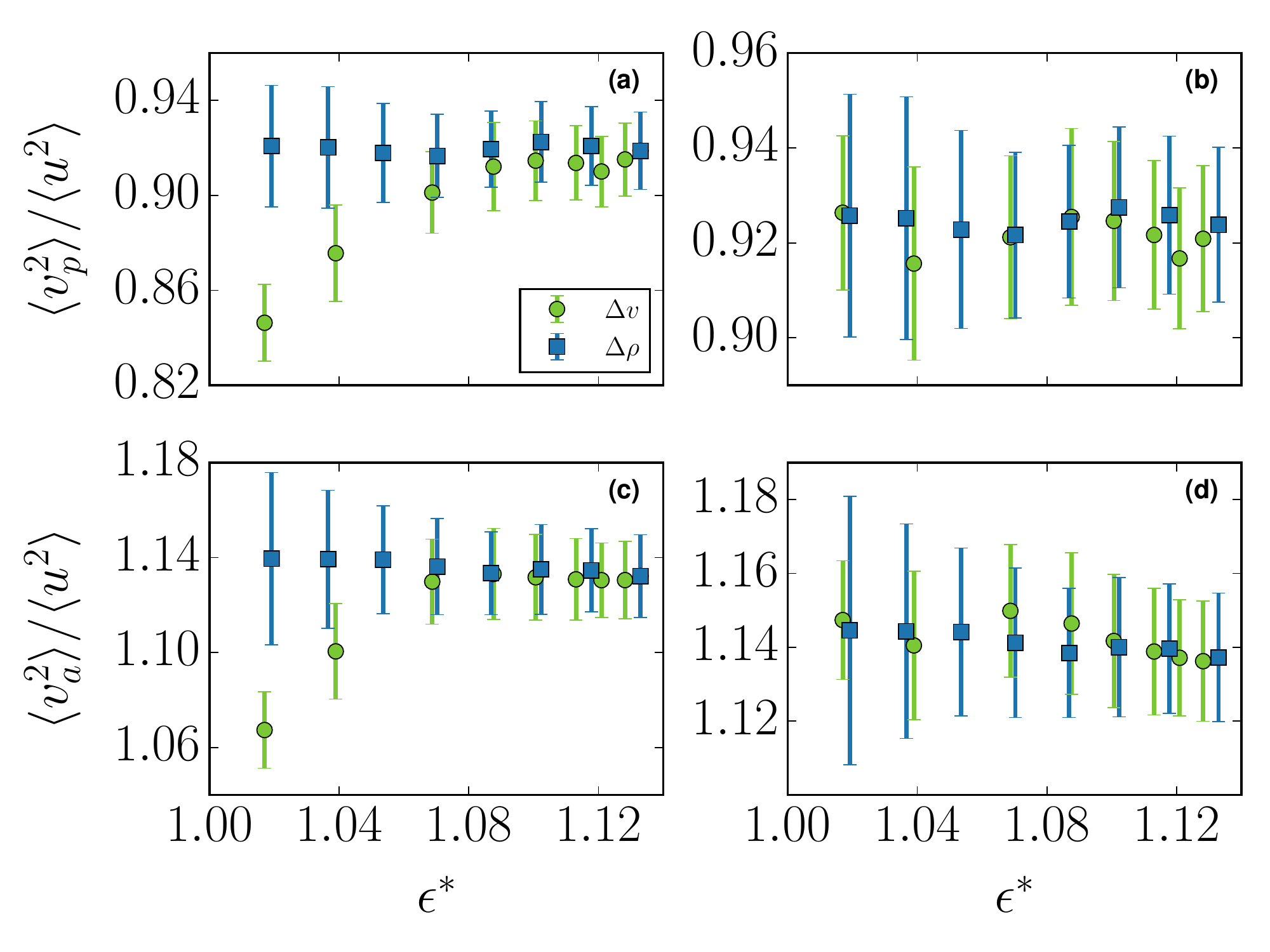}
  \caption{Relative energy density at positions of passive (a) and active (c) matter, as a fraction of the average energy of the fluid, as a function of the effective energy density $\epsilon^*$.  Corrections for the shape of swimmers, due to the effective radius of the tumbling phase, are applied in (b) and (d), showing that $\epsilon^*$ is capable of aligning the curves, as detailed in the main text.  Green circles indicate constant density with varying velocity values, and blue squares indicate constant velocity with varying density values.}
  \label{fgr:relative_energy}
\end{figure}

The fractional energy density at the location of swimmers and passive particles, as granted by the active matter, is shown in Fig. \ref{fgr:relative_energy}.
Parts (a) and (c) give the passive and active fractions, respectively; while $\epsilon^*$ performs well at higher values, it fails to capture the qualitative nature of low-velocity suspensions.  This can be understood in terms of the exclusion radius around a particle.  Since particles experience steric interactions, there is a limit to how close they can be to one another, which limits the potential strength of the fluid field at their centers.  This is why the energy density for passive particles, which have an exclusion radius of $3w$, is below one.  Swimmers have an exclusion radius of only $w$, and, owing to their tendency to attract each other and form groups, they reach values above one on average.  Swimmers and passive particles also attract each other, but the larger exclusion radius has a greater impact on the total.  Note that there is no ``self-interaction'' energy density in these calculations.  To see how this leads to the qualitative change at low velocity values, consider the effect of tumbling.  When a swimmer enters the tumbling phase, it rotates randomly about its center in order to change its orientation for the next run phase.  This causes the particle to have an effective contact radius equal to half its length, rather than half its width (here $l=3w$).  While this is irrelevant during the tumbling phase itself (since the swimmer is not producing a directed flow), this nevertheless increases the minimum distance between particles at the \emph{start} of a run phase.  In the limit of very low swimming velocity, this means that swimmers never reach the minimum contact distance of a run phase, which necessarily reduces the average energy density each particle experiences.  In contrast, at high velocities, the travel time back to the minimum distance is only a small fraction of a run, and barely impacts the results.  Thus, the two curves align at upper velocities, but separate at lower velocities.  We can compensate for this by adding back the ``lost'' energy density, with the following approximation:
\begin{equation}
 \langle \epsilon_l^* \rangle = \frac{\beta (l-w)}{2 v \tau},
\end{equation}
where $\beta$ is the energy density lost between the two exclusion radii, and $(l-w)/2$ is the change in the exclusion radius itself.  When this is divided by the run length $v\tau$, it represents the fraction of time (and thus the impact on the average) that a swimmer spends before it reaches the nearby particle.  Parts (b) and (d) show this correction, which essentially removes the shape factor of the energy density, and realigns the two curves along the single parameter $\epsilon^*$.

\begin{figure}
\centering
  \includegraphics[width=\linewidth]{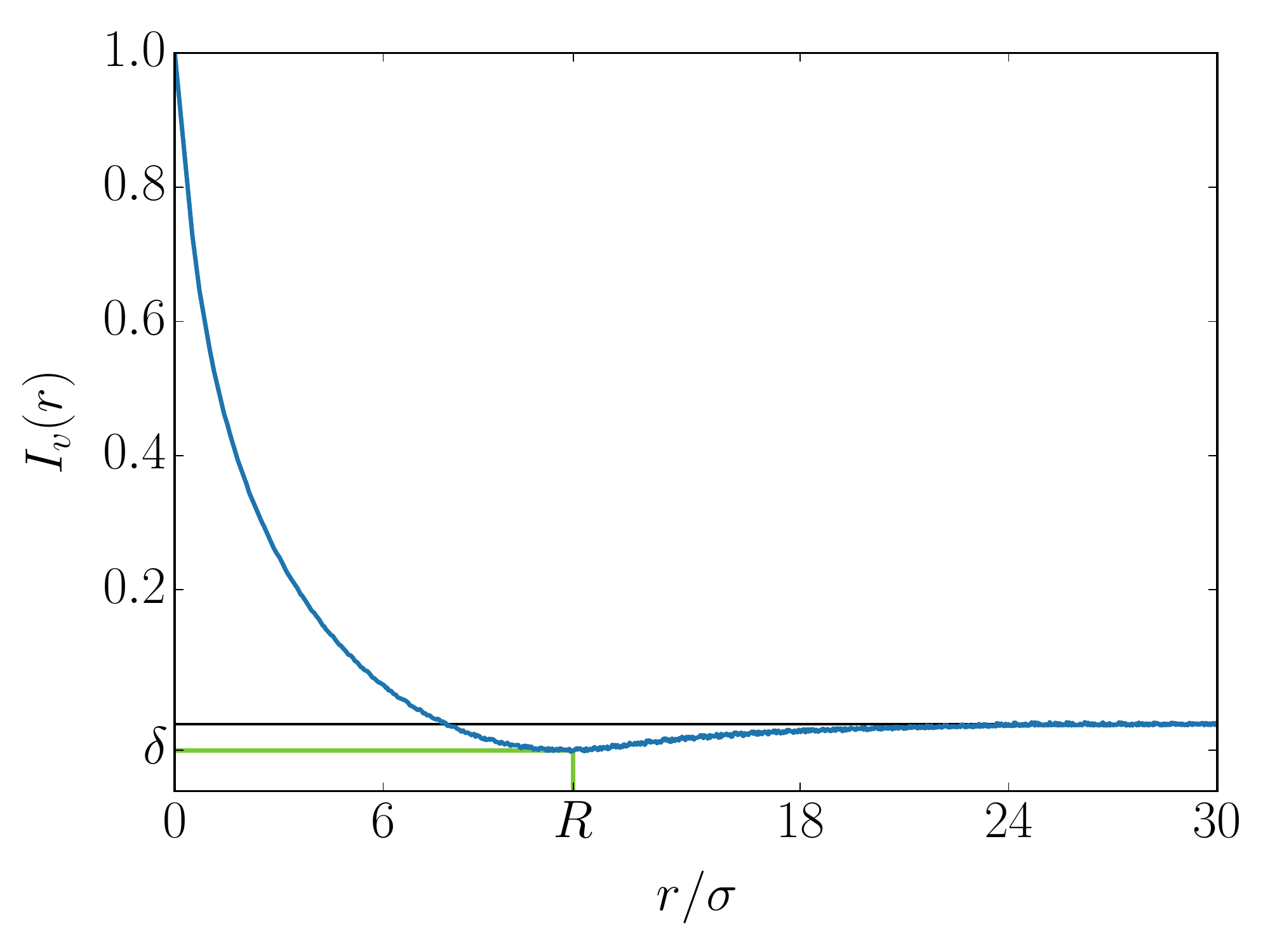}
  \caption{Spatial velocity autocorrelation function $I_v$ as a function of the distance $r$ divided by the interparticle spacing $\sigma$.  The negative correlation $\delta$ indicates the magnitude of long-distance coordination in the fluid, while the point $R$ indicates its size.  The shape of this plot is identical for each simulation, although the values of $\delta$ and $R$ change based on $\epsilon^*$ and $\rho$, respectively.}
  \label{fgr:vacr_example}
\end{figure}

\begin{figure}
\centering
  \includegraphics[width=\linewidth]{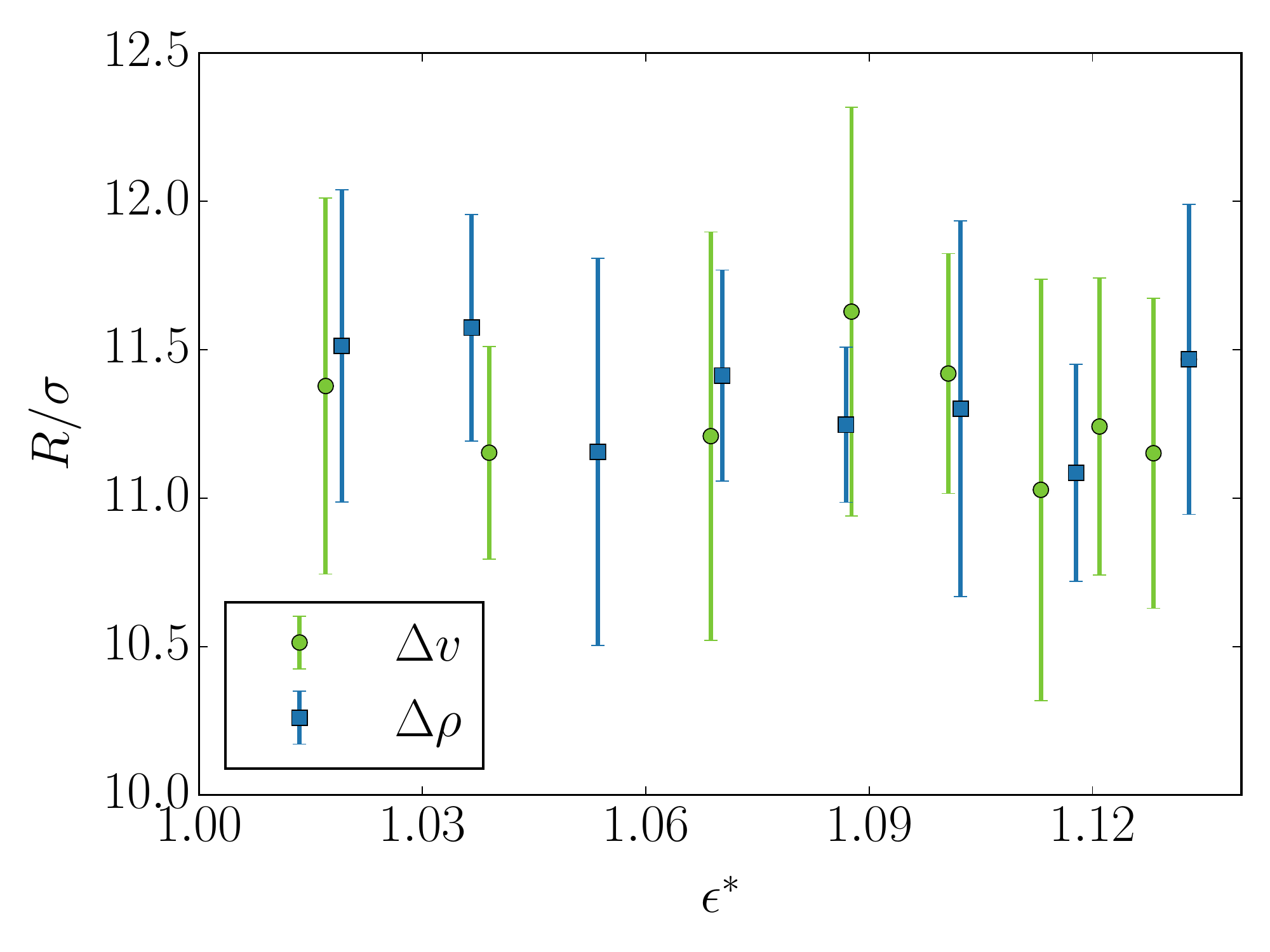}
  \caption{Long-distance coordination length $R$ in units of the interparticle spacing $\sigma$ as a function of the effective energy density $\epsilon^*$.  The coordination scale remains a self-similar constant across all values of $\epsilon^*$ examined.  Green circles indicate constant density with varying velocity values, and blue squares indicate constant velocity with varying density values.}
  \label{fgr:vacr_rmin}
\end{figure}

\begin{figure}
\centering
  \includegraphics[width=\linewidth]{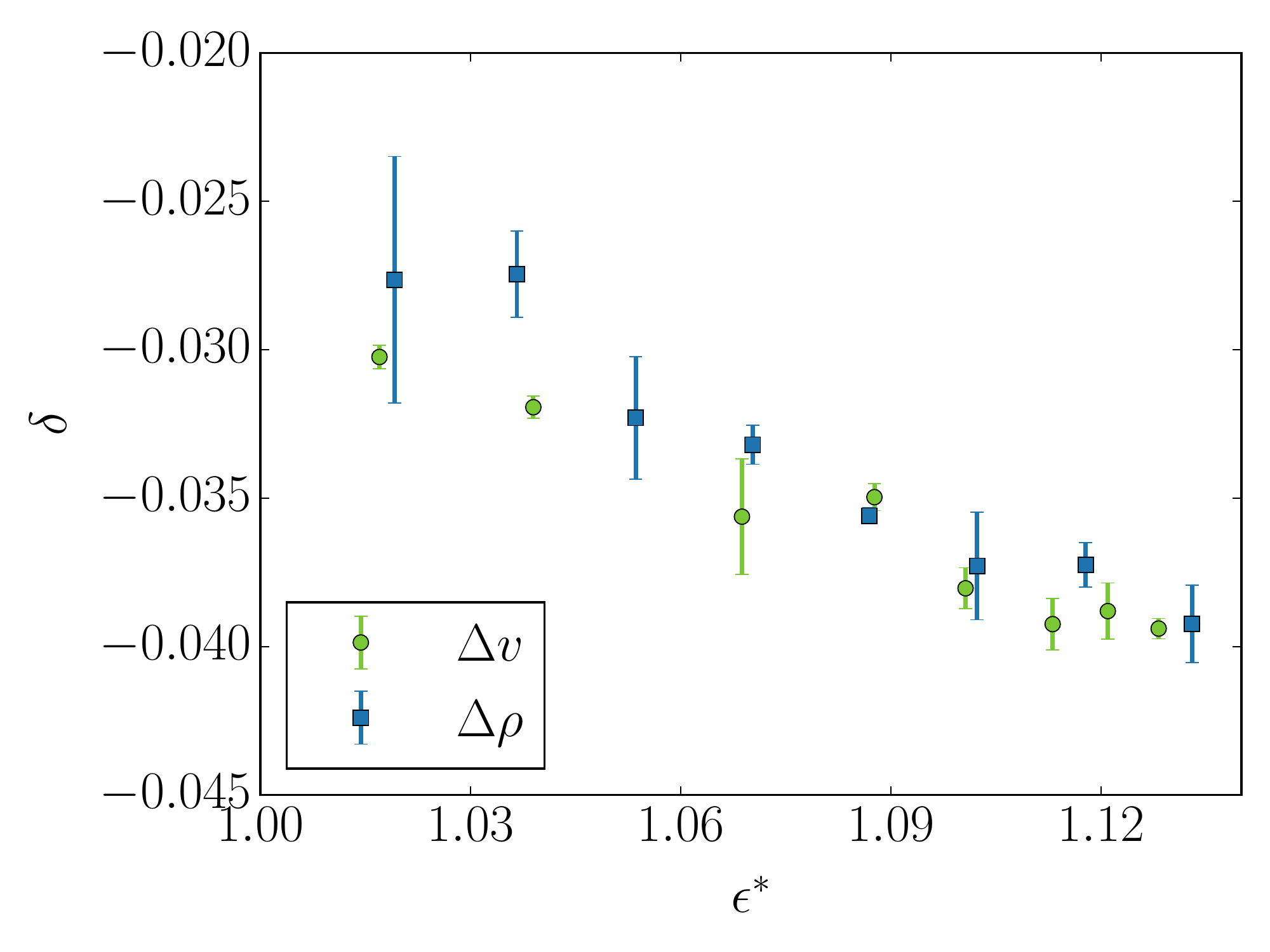}
  \caption{Spatial velocity autocorrelation function depth $\delta$ as a function of the effective energy density $\epsilon^*$.  The data show a roughly linear descrease with $\epsilon^*$, and follows the same line for both parameter changes.  Thus, $\epsilon^*$ is capable of yielding larger scale information in addition to point energy distributions.  Green circles indicate constant density with varying velocity values, and blue squares indicate constant velocity with varying density values.}
  \label{fgr:vacr_vmin}
\end{figure}

While the distribution of energy gives insight into the activity at specific points, it doesn't tell much about the behavior of the system as a whole.  In order to investigate this, we turn to the spatial velocity autocorrelation function, defined by
\begin{equation}
 I_v(\Delta r) = \frac{\langle \mathbf{u}(\mathbf{r},t) \cdot \mathbf{u}(\mathbf{r}+\Delta \mathbf{r},t) \rangle}{\langle u^2 \rangle},
\end{equation}
where the averages are taken over space and time.  An example plot from one simulation, which shares the same shape as any other in this parameter range, is given in Fig. \ref{fgr:vacr_example}.
In such a plot, a negative region indicates long distance coordination, including the formation of swirls.\cite{Kessler2007,Graham2009,Goldstein2013} The location of this minimum is marked by $R$, and the depth by $\delta$.  Since $I_v$ is nondimensionalized by the average fluid energy, $\delta$ indicates the relative strength of the coordination.  $R$ gives its size, in units of the interparticle spacing,
\begin{equation}
 \sigma = \frac{1}{\sqrt{\pi\rho w^2}}.
\end{equation}
At the highest density studied, $\sigma \approx 8$ (eight times the width of a swimmer).  When formatted in this way, $R$ is, remarkably, identical for any value of the density and velocity, as pictured in Fig. \ref{fgr:vacr_rmin}.
This means that, if we zoom in on the system at the right distance so that each one appears to have the same number density, we see the same structure of long-distance coordination.  In other words, this is a self-similar property of the solutions.  While difficult to compare a truly-2D simulation with pseudo-2D experiments, the value of $R \approx 11\sigma$ seems reasonably in line with the experimental work of Cisneros et al.\cite{Kessler2007} The strength of the correlation itself decays roughly linearly with $\epsilon^*$, and is well represented by it, as shown in Fig. \ref{fgr:vacr_vmin}.
That is, the energy of large-scale structures in the fluid does itself correlate with the energy at individual points.  Thus, $\epsilon^*$ is not only useful as a measure of point-averages taken over the whole system, but for measuring the strength of extended structures as well.  This is not particularly surprising when we consider that the energy of any hydrodynamic structure must come from the available energy produced by the swimmers themselves, which is, as has been shown, a function of $\epsilon^*$, rather than $\epsilon$.  In our simulations, we see the onset of many large-scale swirling formations, although they are predominantly transient.  Keep in mind that, at these densities, ``large-scale'' vortices do not include a large number of swimmers in tight polar order along the perimeter, unlike the wide packs of ordered swimmers in dense systems.\cite{Kessler2007,Goldstein2013}  Again, the strength of these swirls is comparable in magnitude to those investigated by Cisneros et al.,\cite{Kessler2007} which is not particularly surprising, even given the difference in parameters and setup, since the value itself does not change significantly.

\begin{figure}
\centering
  \includegraphics[width=\linewidth]{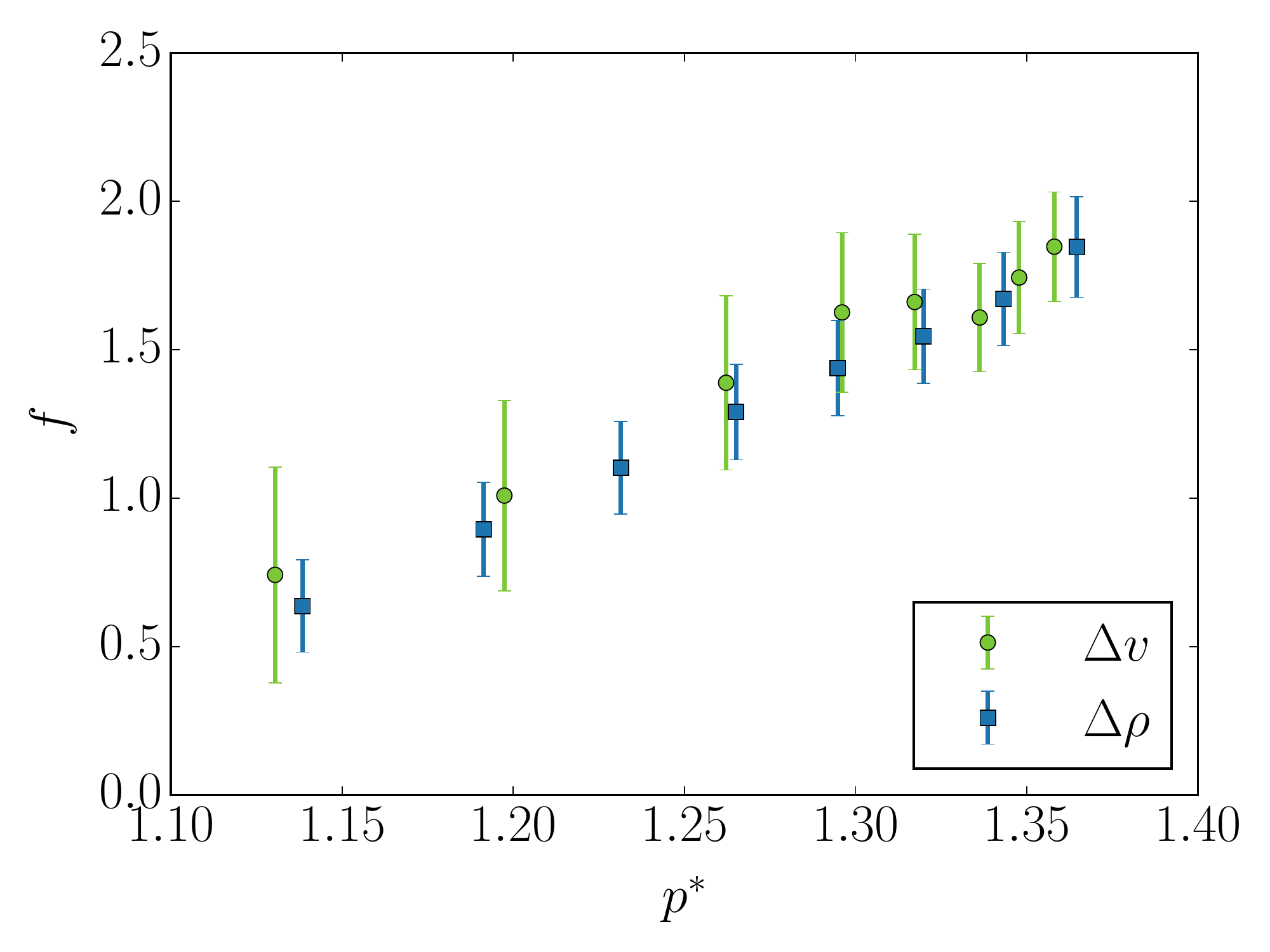}
  \caption{Frequency of passive particle events (per particle, per run time $\tau$) as a function of effective momentum density $p^*$.  This represents all collisional events between passive particles, such as pair formation and deformation and large-scale cluster activity.  The parameter $p^*$ is a simple function of $\epsilon^*$, as given in the text, and aligns the curves along different parameters.  Green circles indicate constant density with varying velocity values, and blue squares indicate constant velocity with varying density values.}
  \label{fgr:event_rate}
\end{figure}

An important property of passive particles in an active bath is the lifetime of their combined structures, such as the pairs formed by effective attractions.\cite{DiLeonardo2011,Garcia2015}  The rate at which particles combine and separate can be defined in terms of an event rate parameter $f$, the number of events per particle per time interval $\tau$, displayed in Fig. \ref{fgr:event_rate}.
Since this parameter is a rate rather than an energy, we use a derived momentum-like quantity for the dependent variable,
\begin{equation}
 p^* = 1 + \sqrt{\epsilon^* - 1},
\end{equation}
which is simply formed by taking the square root of the energy density change as the momentum change.  The event rate has a roughly linear dependence on $p^*$, and the alignment of the two curves is once again excellent, displaying the ability of $\epsilon^*$ to accurately describe the system for rates in addition to energies.  That is, the additional momentum of the fluid field is adequately described with the same clustering method, and determines the rate at which passive particles collide into one another and split apart.  Unsurprisingly, the scale of the event rate is equal to one event per particle, per run duration $\tau$.  Free or paired passive particles are most likely to remain stable on this timescale since anything nearby (already connected or otherwise) is expected to move in a similar fashion.  Larger clusters are more likely to break up since, as can be clearly seen in Fig \ref{fgr:vacr_example}, the correlation of fluid velocity drops substantially within a few $\sigma$.

\begin{figure}
\centering
  \includegraphics[width=\linewidth]{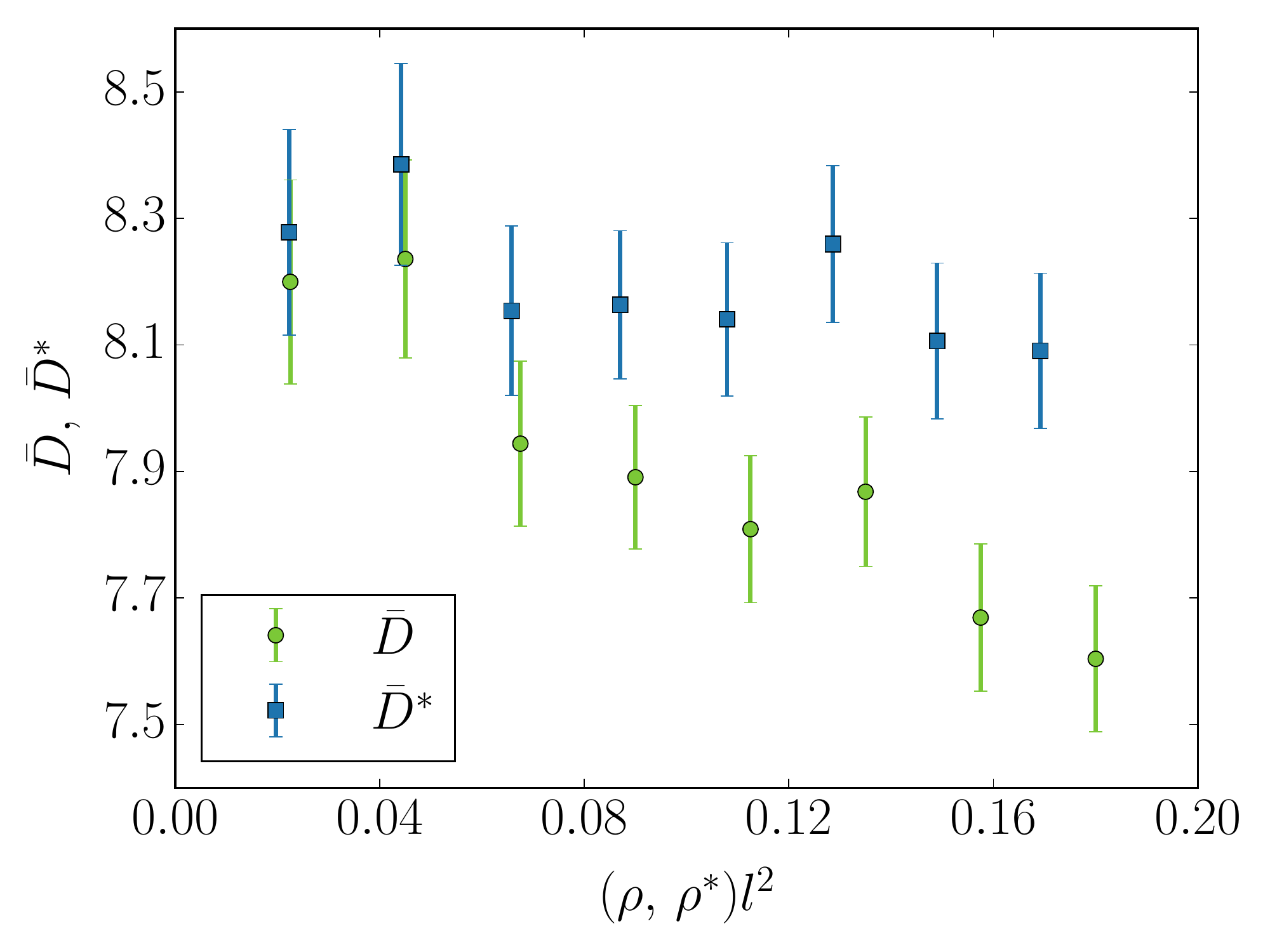}
  \caption{Passive particle dimensionless diffusivity $\bar{D}$ as a function of the active particle density.  Diffusion is impossible to align at low values of velocity and density using $\epsilon^*$ since, at low velocity, the correlation of particle velocity skyrockets.  However, the effective number density $\rho^*$ gives results much closer to the experimental linear relationship between diffusion and density,\cite{Libchaber2000,Arratia2016} which would yield a constant value of $\bar{D}$.  Green circles indicate input density, and blue squares indicate effective density.}
  \label{fgr:diffusion}
\end{figure}

In contast to the above properties, it is difficult for $\epsilon^*$ to capture the qualitative diffusive properties of low-velocity systems.  This is because, as the swimming velocity approaches zero, the spatial configuration of the system becomes a constant, leading to very high velocity correlation.  Obviously, a low-density, high-velocity system will never capture the same physics, so equivalent values of $\epsilon^*$ must give different values.  The essential fact here is that long-term diffusion is more about the speed at which the swimmers move than it is about the strength of the velocity flow coming off of them.  Patteson et al.\cite{Arratia2016} have found that the P\'eclet number, $Pe=vl/D_0$, is a suitable parameter for collapsing different velocity and density values when describing passive particle diffusivities in an active bath, where $D_0$ is the diffusivity of a thermal system without active matter.  In our calculations, in order to specifically examine the active contributions, we have neglected $D_0$, so $Pe=\infty$.  However, the clustering algorithm and the concept of $\epsilon^*$ can still be leveraged to improve upon this relationship.  The diffusion is defined as
\begin{equation}
 D = \frac{\langle (\Delta \mathbf{r})^2 \rangle}{4\Delta t},
\end{equation}
which is simply the mean squared displacement (MSD) on a time interval $\Delta t$.  With active systems, this value changes depending on the time scale (super-diffusive at short times, simple diffusion at long times), so we take the long time limit of the slope of MSD vs. time in order to determine the diffusion.  Following Patteson et al., we also define the dimensionless diffusivity as\cite{Arratia2016}
\begin{equation}
 \bar{D} = \frac{D}{\rho v l^3},
\end{equation}
for 2D.  This quantity should be independent of the density, and depend merely on $v\tau/l$, since we are at high $Pe$.  However, we see in Fig. \ref{fgr:diffusion} a linear decay of $\bar{D}$ with swimmer density for the unaltered value.
When we use the results from clustering to extract the effective density $\rho^*$ from $\epsilon^*$, the agreement improves substantially.  This is because, when we increase the density, some of the added amount is \emph{lost} to attractive interactions, and the variance in the velocity field, over time, becomes lower than anticipated.  Meanwhile, these formations \emph{increase} the energy density, as we have seen above.  Despite the improvement, there is still a slight downward trend in the data, indicating that this simple approximation is not a perfect tool for explaining the density dependence of $D$ at high values of $Pe$. Certainly there must be some impact of the additional fluid velocity granted by groups of swimmers, although, understandably, not in the same fashion as the energy density itself.  As already mentioned, the assumption that a cluster of swimmers acts as a single swimmer at the center would also affect this result, since the spatial distribution of the swimmers is especially important for diffusivity.  A more refined approach to determining the effective density, as pertains to particle diffusion, would therefore yield an even better correction.  Nevertheless, the improvement granted by this simple algorithm is substantial, and shows the potential for it to prove useful even in situations where it is not obviously applicable.

\section{Conclusions}
We have investigated a new parameter, dubbed the effective energy density $\epsilon^*$, and its ability to uniquely predict the properties of hybrid solutions containing both active and passive matter.  It unifies the (coupled) effects of swimmer density and velocity to improve our understanding of how energy is applied to a system through its active elements.  The parameter excels at describing the energy distribution in the fluid, both on average and at the locations of each type of particle.  The energy at particle locations can be understood in terms of the exclusion volume surrounding the particle (a reduction), its interactions with the swimmers (positive for attraction, negative for repulsion), and the available energy within the fluid itself.  Furthermore, it predicts the strength of larger-scale structures within the fluid, such as vortices.  When converted into the momentum-like quantity $p^*$, it can also be used for determining the event rates of a system, including particle collisions and pair formation.  None of these measurements can be adequately explained using the unaltered energy density, as the two parameters have completely different scaling with density and velocity.  While it cannot describe the qualitative roles of velocity and density in passive particle diffusivity, the derived effective density, $\rho^*$, allows the results to reflect the experimental predictions\cite{Libchaber2000,Arratia2016} much better that the raw swimmer density itself.  Despite this success, the clustering algorithm is limited in its efficacy to systems below the threshold of large-scale collective motion, such as turbulence\cite{Goldstein2013}, and when considering properties that depend strongly on the distance the swimmers travel, or the spatial extension of groups of swimmers, rather than the hydrodynamic field they create specifically.  The spatial separation effect could be compensated for by removing the assumption that clusters act as point particles with a net field based on the particle count.  It would be interesting to see if enhancements to the algorithm used to define $\epsilon^*$ could allow its predictions to extend to the dynamics of highly coordinated swimmer motion in dense systems, and could be explored in future work on the matter.

\section{Acknowledgements}
This work was funded by the National Institutes of Health grant GM086801, the National Science Foundation grant MCB-1050966, and the U.S. Department of Energy through the LANL/LDRD Program.  Computations were performed on the Extreme Science and Engineering Discovery Environment (XSEDE).





\providecommand*{\mcitethebibliography}{\thebibliography}
\csname @ifundefined\endcsname{endmcitethebibliography}
{\let\endmcitethebibliography\endthebibliography}{}

\end{document}